\documentclass[letterpaper,12pt]{article}
\usepackage[margin=1in]{geometry}
\usepackage{verbatim}
\usepackage{amsmath}
\usepackage{amssymb}
\usepackage{graphicx}
\usepackage{amsthm}

\font\dsfont=dsrom12

\def\R{{\hbox{\dsfont R}}}

\def\epsilon{\varepsilon}

\def\P{\mathcal{P}}

\def\phi{\varphi}

\def\0s{{\bf 0}}

\def\Prob{{\bf P}}

\newtheorem{theorem}{Theorem}[section]

\newtheorem{corollary}[theorem]{Corollary}

\newtheorem{observe}[theorem]{Observation}
\newtheorem{remark1}[theorem]{Remark}

\newenvironment{remark}{\begin{remark1} \rm}{\end{remark1}}

\title{Statistical tests for whether a given set
       of independent, identically distributed draws
       does not come from a specified probability density}
\author{Mark Tygert}
\date{June 3, 2010\\\ \\Dedicated to the memory of Sam Roweis}

\begin{document}

\maketitle

\begin{abstract}
We discuss several tests for whether a given set
of independent and identically distributed (i.i.d.)\ draws
does not come from a specified probability density function.
The most commonly used are Kolmogorov-Smirnov tests,
particularly Kuiper's variant,
which focus on discrepancies between the cumulative distribution function
for the specified probability density
and the empirical cumulative distribution function
for the given set of i.i.d.\ draws.
Unfortunately, variations in the probability density function
often get smoothed over in the cumulative distribution function,
making it difficult to detect discrepancies in regions
where the probability density is small
in comparison with its values in surrounding regions.
We discuss tests without this deficiency, complementing the classical methods.
The tests of the present paper are based on the plain fact
that it is unlikely to draw a random number whose probability is small,
provided that the draw is taken from the same distribution used
in calculating the probability
(thus, if we draw a random number whose probability is small,
then we can be confident that we did not draw the number
from the same distribution used in calculating the probability).

\medskip

\end{abstract}

\section{Introduction}

A basic task in statistics is to ascertain
whether a given set of independent and identically distributed (i.i.d.)\ draws
$X_1$,~$X_2$, \dots, $X_{n-1}$,~$X_n$ does not come
from a distribution with a specified probability density function $p$
(the null hypothesis is that $X_1$,~$X_2$, \dots, $X_{n-1}$,~$X_n$
do in fact come from the specified $p$).
In the present paper, we consider the case
when $X_1$,~$X_2$, \dots, $X_{n-1}$,~$X_n$ are real valued.
In this case, the most commonly used approach is due to Kolmogorov and Smirnov
(with a popular modification by Kuiper);
see, for example, Sections~14.3.3 and~14.3.4
of~\cite{press-teukolsky-vetterling-flannery}, \cite{stephens2},
\cite{stephens1}, or Section~\ref{theory} below.

The Kolmogorov-Smirnov approach considers the size of the discrepancy
between the cumulative distribution function for $p$
and the empirical cumulative distribution function
defined by $X_1$,~$X_2$, \dots, $X_{n-1}$,~$X_n$
(see, for example, Sections~\ref{notation} and~\ref{theory} below
for definitions of cumulative distribution functions
and empirical cumulative distribution functions).
If the i.i.d.\ draws $X_1$,~$X_2$, \dots, $X_{n-1}$,~$X_n$ used to form
the empirical cumulative distribution function are taken
from the probability density function $p$ used in the Kolmogorov-Smirnov test,
then the discrepancy is small.
Thus, if the discrepancy is large,
then we can be confident that $X_1$,~$X_2$, \dots, $X_{n-1}$,~$X_n$
do not come from a distribution with probability density function $p$.

However, the size of the discrepancy between
the cumulative distribution function for $p$
and the empirical cumulative distribution function
constructed from the i.i.d.\ draws $X_1$,~$X_2$, \dots, $X_{n-1}$,~$X_n$
does not always signal that $X_1$,~$X_2$, \dots, $X_{n-1}$,~$X_n$
do not arise from a distribution
with the specified probability density function $p$,
even when $X_1$,~$X_2$, \dots, $X_{n-1}$,~$X_n$ do not in fact arise from $p$.
In some cases, $n$ has to be absurdly large for the discrepancy
to be significant. It is easy to see why:

The cumulative distribution function is an indefinite integral
of the probability density function $p$.
Therefore, the cumulative distribution function is a smoothed version
of the probability density function;
focusing on the cumulative distribution function rather than $p$ itself
makes it harder to detect discrepancies in regions where $p$ is small
in comparison with its values in surrounding regions.
For example, consider the probability density function $p$
depicted in Figure~\ref{bimodal_fig} below 
(a ``tent'' with a narrow triangle removed at its apex)
and the probability density function $q$
depicted in Figure~\ref{unimodal_fig} below
(nearly the same ``tent,'' but with the narrow triangle intact, not removed).
The cumulative distribution functions for $p$ and $q$ are very similar,
so tests of the classical Kolmogorov-Smirnov type have trouble
signaling that i.i.d.\ draws taken from $q$ are actually not taken from $p$.
Section~14.3.4 of~\cite{press-teukolsky-vetterling-flannery} highlights
this problem and a strategy for its solution,
hence motivating us to write the present article.

We propose to supplement tests of the classical Kolmogorov-Smirnov type
with tests for whether any of the values
$p(X_1)$,~$p(X_2)$, \dots, $p(X_{n-1})$,~$p(X_n)$ 
is small. If any of these values is small, then we can be confident
that the i.i.d.\ draws $X_1$,~$X_2$, \dots, $X_{n-1}$,~$X_n$ did not
arise from the probability density function $p$.
Theorem~\ref{newdist_thm} below formalizes the notion of any
of $p(X_1)$,~$p(X_2)$, \dots, $p(X_{n-1})$,~$p(X_n)$ being small.
We also propose another complementary test, which amounts
to using the Kolmogorov-Smirnov approach after ``rearranging''
the probability density function $p$ so that it is nondecreasing
on the shortest interval outside which it vanishes
(see Remark~\ref{rearrangement} and formula~(\ref{Kuiper2}) below).

For descriptions of other generalizations of and alternatives to
the Kolmogorov-Smirnov approach
(concerning issues distinct from those treated in the present paper),
see, for example,
Sections~14.3.3 and~14.3.4 of~\cite{press-teukolsky-vetterling-flannery},
\cite{barron}, \cite{bickel-rosenblatt}, \cite{fan}, \cite{hollander-wolfe},
\cite{inglot-ledwina}, \cite{khamis}, \cite{rayner-thas-best},
\cite{reschenhofer}, \cite{simonoff2}, \cite{stephens2}, \cite{wasserman},
and their compilations of references.
For a more general approach, based on customizing statistical tests
for problem-specific families of alternative hypotheses,
see~\cite{bickel-ritov-stoker}.
Below, we compare the test statistics of the present article
with one of the most commonly used test statistics
of the Kolmogorov-Smirnov type, namely Kuiper's
(see, for example, \cite{stephens2}, \cite{stephens1},
or Section~\ref{theory} below).
We recommend using the test statistics of the present paper
in conjunction with the Kuiper statistic, to be conservative,
as all these statistics complement each other,
helping compensate for their inevitable deficiencies.

There are at least two canonical applications.
First, the tests of the present article can be suitable for checking
for malfunctions with and bugs in computer codes that are supposed
to generate pseudorandom i.i.d.\ draws
from specified probability density functions
(especially the complicated ones encountered frequently in practice).
Good software engineering requires such independent tests
for helping validate that computer codes produce correct results
(of course, such validations do not obviate careful,
structured programming, but are instead complementary).
Second, many theories from physics and physical chemistry
predict (often {\it a priori}) the probability density functions
from which experiments are supposed to be taking i.i.d.\ draws.
The tests of the present paper can be suitable for ruling out
erroneous theories of this type, on the basis of experimental data.
Moreover, there are undoubtedly many other potential applications,
in addition to these two.

For definitions of the notation used throughout, see Section~\ref{notation}.
Section~\ref{theory} introduces several statistical tests.
Section~\ref{numerical} illustrates the power of the statistical tests
via some numerical examples.
Section~\ref{conclusions} draws several conclusions and proposes directions
for further work.

\section{Notation}
\label{notation}

In this section, we set notation used throughout the present paper.

We use $\Prob$ to take the probability of an event.
We say that $p$ is a probability density function
to mean that $p$ is a (Lebesgue-measurable) function from $\R$ to $[0,\infty)$
such that the integral of $p$ over $\R$ is 1.

The cumulative distribution function $P$
for a probability density function $p$ is
\begin{equation}
\label{cumulative}
P(x) = \int_{y \le x} p(y) \, dy
\end{equation}
for any real number $x$.
If $X$ is a random variable distributed according to $p$,
then $P(x)$ is just the probability that $X \le x$.
Therefore, if $X$ is a random variable distributed according to $p$,
then the cumulative distribution function $\P$ for $p(X)$ is
\begin{equation}
\label{rearranged}
\P(x) = \int_{p(y) \le x} p(y) \, dy,
\end{equation}
the probability that $p(X) \le x$.

For reference, we summarize our (reasonably standard) notational conventions
in Table~\ref{notations}.

\begin{table}
\caption{Notational conventions}
\label{notations}
\vspace{1em}
\begin{tabular*}{\textwidth}{@{\extracolsep{\fill}}lll}
{\it mathematical object} & {\it typeface} & {\it example} \\\hline
probability density function & italic lowercase & $p(x)$ \\
cumulative distribution function defined in~(\ref{cumulative})
 & italic uppercase & $P(x)$ \\
distribution function defined in~(\ref{rearranged}) & script uppercase
 & $\P(x)$ \\
taking the probability of an event & bold uppercase
 & $\Prob\bigl\{X \le x\bigr\}$
\end{tabular*}
\end{table}

\begin{remark}
\label{rearrangement}
The ``nonincreasing rearrangement'' (or nondecreasing rearrangement)
of a probability density function
(see, for example, Section V.3 of~\cite{stein-weiss})
clarifies the meaning of the distribution function $\P$
defined in~(\ref{rearranged}).
With $P$ defined in~(\ref{cumulative}) and $\P$ defined in~(\ref{rearranged}),
$\P(p(x)) = P(x)$ for any real number $x$ in the shortest interval
outside which the probability density function $p$ vanishes,
as long as $p$ is increasing on that shortest interval.
\end{remark}

\section{Test statistics}
\label{theory}

In this section, we introduce several statistical tests.

One test of whether i.i.d.\ draws $X_1$,~$X_2$, \dots, $X_{n-1}$,~$X_n$
do not arise from a specified probability density function $p$
is the Kolmogorov-Smirnov test (or Kuiper's often preferable variation).
If $X$ is a random variable distributed according to $p$,
then another test is to use the Kolmogorov-Smirnov or Kuiper test
for the random variable $p(X)$, whose cumulative distribution function
is $\P$ in~(\ref{rearranged}).
The test statistic for the original Kuiper test is
\begin{equation}
\label{Kuiper1}
U = \left( \sqrt{n} \, \sup_{-\infty < x < \infty} P(x) - \hat{P}(x) \right)
  - \left( \sqrt{n} \, \inf_{-\infty < x < \infty} P(x) - \hat{P}(x) \right),
\end{equation}
where $\hat{P}(x)$ is the empirical cumulative distribution function
--- the number of $k$ such that $X_k \le x$, divided by $n$.
The test statistic for the Kuiper test for $p(X)$ is therefore
\begin{equation}
\label{Kuiper2}
V = \left( \sqrt{n} \, \sup_{0 \le x < \infty} \P(x) - \hat{\P}(x) \right)
  - \left( \sqrt{n} \, \inf_{0 \le x < \infty} \P(x) - \hat{\P}(x) \right),
\end{equation}
where $\hat{\P}(x)$ is the number of $k$ such that $p(X_k) \le x$,
divided by $n$.
Remark~\ref{rearrangement} above and Remark~\ref{other_norms} below
provide some motivation for using~$V$, beyond its being a natural variation
on $U$.

The rationale for using statistics such as $U$ and $V$
is the following theorem, corollary, and the ensuing discussion
(see, for example, Sections~14.3.3 and~14.3.4
of~\cite{press-teukolsky-vetterling-flannery}, \cite{stephens1},
or~\cite{stephens2} for proofs and details).

\begin{theorem}
\label{uni1}
Suppose that $p$ is a probability density function,
$X$ is a random variable distributed according to $p$,
and $P$ is the cumulative distribution function for $X$
from~(\ref{cumulative}).
Then, the distribution of $P(X)$ is the uniform distribution over $[0,1]$.
\end{theorem}

\begin{corollary}
\label{uni2}
Suppose that $p$ is a probability density function,
$X$ is a random variable distributed according to $p$,
and $\P$ is the cumulative distribution function for $p(X)$
from~(\ref{rearranged}).
Then, the cumulative distribution function of $\P(p(X))$
is less than or equal to the cumulative distribution function
of the uniform distribution over $[0,1]$.
Moreover, the distribution of $\P(p(X))$
is the uniform distribution over $[0,1]$ if $\P$ is a continuous function
($\P$ is a continuous function when, for every nonnegative real number $y$,
the probability that $p(X) = y$ is 0).
\end{corollary}

Theorem~\ref{uni1} generalizes to the fact that,
if the i.i.d.\ draws $X_1$,~$X_2$, \dots, $X_{n-1}$,~$X_n$
arise from the probability density function $p$ involved
in the definition of $U$ in~(\ref{Kuiper1}),
then the distribution of $U$ does not depend on $p$;
the distribution of $U$ is the same for any $p$.
With high probability, $U$ is not much greater than 1
when the i.i.d.\ draws $X_1$,~$X_2$, \dots, $X_{n-1}$,~$X_n$
used in the definition of $U$ in~(\ref{Kuiper1})
are taken from the distribution whose probability density function $p$
and cumulative distribution function $P$ are used in the definition of $U$.
Therefore, if the statistic $U$ that we compute
turns out to be substantially greater than 1,
then we can have high confidence that the i.i.d.\ draws
$X_1$,~$X_2$, \dots, $X_{n-1}$,~$X_n$ were not taken from the distribution
whose probability density function $p$ and cumulative distribution function $P$
were used in the definition of $U$.
Similarly, if $V$ defined in~(\ref{Kuiper2}) turns out to be substantially
greater than 1, then we can have high confidence that the i.i.d.\ draws
$X_1$,~$X_2$, \dots, $X_{n-1}$,~$X_n$ were not taken from the distribution
whose probability density function $p$ and distribution function $\P$
were used in the definition of $V$.
For details, see, for example, Sections~14.3.3 and~14.3.4
of~\cite{press-teukolsky-vetterling-flannery}, \cite{stephens1},
or~\cite{stephens2}.

A third test statistic is
\begin{equation}
\label{simple}
W = n \, \min_{1 \le k \le n} \P(p(X_k)).
\end{equation}
The following theorem and ensuing discussion characterize $W$
and its applications.

\begin{theorem}
\label{newdist_thm}
Suppose that $p$ is a probability density function, $n$ is a positive integer,
$X_1$,~$X_2$, \dots, $X_{n-1}$,~$X_n$ are i.i.d.\ random variables
each distributed according to $p$,
$\P$ is the cumulative distribution function for $p(X_1)$
from~(\ref{rearranged}),
and $W$ is the random variable defined in~(\ref{simple}).
Then,
\begin{equation}
\label{newdist}
\Prob\bigl\{ W \le x \bigr\} \le 1 - \left(1-\frac{x}{n}\right)^n
\end{equation}
for any $x \in [0,n]$.
\end{theorem}

\begin{proof}
It follows from~(\ref{simple}) that
\begin{equation}
\label{first}
\Prob\bigl\{ W > nx \bigr\}
= \Prob\bigl\{ \P(p(X_1)) > x \hbox{ and } \P(p(X_2)) > x
               \hbox{ and \dots\ and } \P(p(X_n)) > x \bigr\}
\end{equation}
for any $x \in [0,1]$.
It follows from the independence of $X_1$,~$X_2$, \dots, $X_{n-1}$,~$X_n$ that
\begin{equation}
\label{second}
\Prob\bigl\{ \P(p(X_1)) > x \hbox{ and } \P(p(X_2)) > x
             \hbox{ and \dots\ and } \P(p(X_n)) > x \bigr\}
= \prod_{k=1}^n \Prob\bigl\{ \P(p(X_k)) > x \bigr\}
\end{equation}
for any $x \in [0,1]$.
It follows from Corollary~\ref{uni2} that
\begin{equation}
\label{third}
\Prob\bigl\{ \P(p(X_k)) > x \big\} \ge 1-x
\end{equation}
for any $x \in [0,1]$ and $k = 1$,~$2$, \dots, $n-1$,~$n$.
Combining~(\ref{first}), (\ref{second}), and~(\ref{third})
yields~(\ref{newdist}).
\end{proof}

For any positive real number $\alpha < 1/2$, we define
\begin{equation}
\label{threshold}
x_\alpha = n-n(1-\alpha)^{1/n};
\end{equation}
if $W \le x_\alpha$, then due to~(\ref{newdist})
we can have at least $[100(1-\alpha)]\%$ confidence that
the i.i.d.\ draws $X_1$,~$X_2$, \dots, $X_{n-1}$,~$X_n$ do not arise from $p$.
It follows from~(\ref{threshold}) that
\begin{equation}
\label{tight_bounds}
\alpha \le x_\alpha < -\ln(1-\alpha)
         = \alpha + \alpha^2/2 + \alpha^3/3 + \alpha^4/4 + \ldots
         < \alpha + \alpha^2,
\end{equation}
with $x_\alpha = \alpha$ for $n = 1$,
and $\lim_{n \to \infty} x_\alpha = -\ln(1-\alpha)$. Therefore,
if $W \le \alpha$, then we have at least $[100(1-\alpha)]\%$ confidence that
the i.i.d.\ draws $X_1$,~$X_2$, \dots, $X_{n-1}$,~$X_n$ do not arise from $p$.
Taking $\alpha=.01$, for example, we have at least 99\% confidence that
the i.i.d.\ draws $X_1$,~$X_2$, \dots, $X_{n-1}$,~$X_n$ do not arise from $p$,
if $W \le .01$.

In short, for any positive real number $\alpha < 1/2$,
if the statistic $W$ defined in~(\ref{simple}) is at most $\alpha$,
then we have at least $[100(1-\alpha)]\%$ confidence that
the i.i.d.\ draws $X_1$,~$X_2$, \dots, $X_{n-1}$,~$X_n$ do not arise
from the probability density function $p$ used in~(\ref{simple}).
If however $W$ is greater than $\alpha + \alpha^2$,
then~(\ref{newdist}) provides no basis for claiming
with at least $[100(1-\alpha)]\%$ confidence that
the i.i.d.\ draws $X_1$,~$X_2$, \dots, $X_{n-1}$,~$X_n$ do not arise
from the probability density function $p$ used in~(\ref{simple}).

\begin{remark}
\label{interpretation}
If $W$ defined in~(\ref{simple}) is at most 1,
then we can have at least $[100(1-W)]\%$ confidence that
the i.i.d.\ draws $X_1$,~$X_2$, \dots, $X_{n-1}$,~$X_n$ do not arise
from the probability density function $p$ used in~(\ref{simple}).
\end{remark}

\begin{remark}
Using $W$ defined in~(\ref{simple}) along with the upper bound~(\ref{newdist})
is optimal when the probability density function $p$
takes on only finitely many values, or when $p$ has the property that,
for every nonnegative real number $y$, the probability is~0
that $p(X) = y$, where $X$ is a random variable distributed according to $p$.
In both cases, the inequality~(\ref{newdist}) becomes the equality
\begin{equation}
\Prob\bigl\{ W \le n \; \P(p(x)) \bigr\} = 1 - \Bigl(1-\P(p(x))\Bigr)^n
\end{equation}
for any $x \in \R$.
\end{remark}

\begin{remark}
\label{other_norms}
When the statistic $W$ defined in~(\ref{simple}) is not powerful enough
to discriminate between two particular distributions,
then a natural alternative is the average
\begin{equation}
\tilde{W} = \frac{1}{n} \sum_{1 \le k \le n} \P(p(X_k)).
\end{equation}
The Kuiper test statistic $V$ defined in~(\ref{Kuiper2})
is a more refined version of this alternative,
and we recommend using $V$ instead of $\tilde{W}$,
in conjunction with the use of $W$ and $U$ defined in~(\ref{Kuiper1}).
We could also consider more general averages of the form
\begin{equation}
f \left( \frac{1}{n} \, \sum_{1 \le k \le n} g\Bigl(\P(p(X_k))\Bigr) \right),
\end{equation}
where $f$ and $g$ are functions;
obvious candidates include $f(x) = \exp(x)$ and $g(x) = \ln(x)$,
and $f(x) = 1-x^{1/q}$ and $g(x) = (1-x)^q$, with $q \in (1,\infty)$.
\end{remark}

\begin{remark}
To clarify further, let us consider the case $n = 1$.
If we are given a probability density function $p$ and a draw $X$
(not necessarily from $p$) such that
$\P(p(X))$ is small, where $\P$ is defined in~(\ref{rearranged}) for $p$,
then why can we be confident that $X$ was not drawn from a distribution
with probability density function $p$?
Well, if $\P(p(X))$ is small, then the likelihood of drawing $X$
from a distribution with probability density function $p$ is small,
in the sense that we would be at least as confident that any draw $Y$ 
satisfying $p(Y) \le p(X)$ does not arise from a distribution
with probability density function $p$,
and the probability under $p$ of all such draws is just $\P(p(X))$,
which is small (by assumption).
\end{remark}

\section{Numerical examples}
\label{numerical}

In this section, we illustrate the effectiveness of the test statistics
of the present paper via several numerical experiments.
For each experiment, we compute the statistics $U$, $V$, and $W$
defined in~(\ref{Kuiper1}), (\ref{Kuiper2}), and~(\ref{simple})
for two sets of i.i.d.\ draws, first for i.i.d.\ draws
$X_1$,~$X_2$, \dots, $X_{n-1}$,~$X_n$ taken
from the distribution whose probability density function $p$,
cumulative distribution function $P$, and distribution function $\P$
are used in the definitions of $U$, $V$, and $W$
in~(\ref{Kuiper1}), (\ref{Kuiper2}), and~(\ref{simple}),
and second for i.i.d.\ draws $X_1$,~$X_2$, \dots, $X_{n-1}$,~$X_n$
taken from a different distribution.

The test statistics $U$ and $V$ defined in~(\ref{Kuiper1}) and~(\ref{Kuiper2})
are the same, except that $U$ concerns a random variable $X$
drawn from a probability density function $p$, while $V$ concerns $p(X)$.
We can directly compare the values of $U$ and $V$
for various distributions in order to gauge
their relative discriminative powers.
Ideally, $U$ and $V$ should not be much greater than 1
when the i.i.d.\ draws $X_1$,~$X_2$, \dots, $X_{n-1}$,~$X_n$
used in the definitions of $U$ and $V$ in~(\ref{Kuiper1}) and~(\ref{Kuiper2})
are taken from the distribution whose probability density function $p$,
cumulative distribution function $P$, and distribution function $\P$
are used in the definitions of $U$ and $V$;
$U$ and $V$ should be substantially greater than 1
when the i.i.d.\ draws $X_1$,~$X_2$, \dots, $X_{n-1}$,~$X_n$
are taken from a different distribution,
to signal the difference between the common distribution
of each of $X_1$,~$X_2$, \dots, $X_{n-1}$,~$X_n$
and the distribution whose probability density function $p$,
cumulative distribution function $P$, and distribution function $\P$
are used in the definitions of $U$ and $V$.

For details concerning the interpretation of and significance levels
for the Kuiper test statistics $U$ and $V$ defined in~(\ref{Kuiper1})
and~(\ref{Kuiper2}), see Sections~14.3.3 and~14.3.4
of~\cite{press-teukolsky-vetterling-flannery},
\cite{stephens2}, or~\cite{stephens1}; both one- and two-tailed
hypothesis tests are available, for any finite number $n$
of draws $X_1$,~$X_2$, \dots, $X_{n-1}$,~$X_n$,
and also in the limit of large $n$.
In short, if $X_1$,~$X_2$, \dots, $X_{n-1}$,~$X_n$ are i.i.d.\ random variables
drawn according to a continuous cumulative distribution function $P$,
then the complementary cumulative distribution function of $U$
defined in~(\ref{Kuiper1})
for the same cumulative distribution function $P$ has an upper tail
that decays nearly as fast as the complementary error function.
Although the details are complicated (varying with $n$ and with the form ---
one-tailed or two-tailed --- of the hypothesis test), the probability that
$U$ is greater than 2 is at most 1\%
when $X_1$,~$X_2$, \dots, $X_{n-1}$,~$X_n$ used in~(\ref{Kuiper1})
are drawn according to the same cumulative distribution function $P$ as used
in~(\ref{Kuiper1}).

As described in Remark~\ref{interpretation}, the interpretation
of the test statistic $W$ defined in~(\ref{simple}) is simple:
If $W$ defined in~(\ref{simple}) is at most 1,
then we can have at least $[100(1-W)]\%$ confidence that
the i.i.d.\ draws $X_1$,~$X_2$, \dots, $X_{n-1}$,~$X_n$ do not arise
from the probability density function $p$ used in~(\ref{simple}).

Tables~\ref{saw_tab}--\ref{smooth_tab} display numerical results
for the examples described in the subsections below.
The following list describes the headings of the tables:

\begin{itemize}
\item $n$ is the number of i.i.d.\ draws $X_1$,~$X_2$, \dots, $X_{n-1}$,~$X_n$
      taken to form the statistics $U$, $V$, and $W$ defined
      in~(\ref{Kuiper1}), (\ref{Kuiper2}), and~(\ref{simple}).
\item $U_0$ is the statistic $U$ defined in~(\ref{Kuiper1}),
      with the $X_1$,~$X_2$, \dots, $X_{n-1}$,~$X_n$ defining $\hat{P}$
      in~(\ref{Kuiper1}) drawn from a distribution
      with the same cumulative distribution function $P$ as used
      in~(\ref{Kuiper1}). Ideally, $U_0$ should be small,
      not much larger than 1.
\item $U_1$ is the statistic $U$ defined in~(\ref{Kuiper1}),
      with the $X_1$,~$X_2$, \dots, $X_{n-1}$,~$X_n$ defining $\hat{P}$
      in~(\ref{Kuiper1}) drawn from a distribution
      with a cumulative distribution function that is different from $P$ used
      in~(\ref{Kuiper1}). Ideally, $U_1$ should be large,
      substantially greater than 1, to signal the difference between
      the common distribution of each of $X_1$,~$X_2$, \dots, $X_{n-1}$,~$X_n$
      and the distribution with the cumulative distribution function $P$
      used in~(\ref{Kuiper1}).
      The numbers in parentheses in the tables indicate the order of magnitude
      of the significance level for rejecting the null hypothesis,
      that is, for asserting that the draws
      $X_1$,~$X_2$, \dots, $X_{n-1}$,~$X_n$ do not arise from $P$.
\item $V_0$ is the statistic $V$ defined in~(\ref{Kuiper2}),
      with the $X_1$,~$X_2$, \dots, $X_{n-1}$,~$X_n$ defining $\hat{\P}$
      in~(\ref{Kuiper2}) drawn from a distribution
      with the same probability density function $p$ used for $\hat{\P}$
      and for $\P$ in~(\ref{Kuiper2}). Ideally, $V_0$ should be small,
      not much larger than 1.
\item $V_1$ is the statistic $V$ defined in~(\ref{Kuiper2}),
      with the $X_1$,~$X_2$, \dots, $X_{n-1}$,~$X_n$ defining $\hat{\P}$
      in~(\ref{Kuiper2}) drawn from a distribution that is different
      from the distribution with the probability density function $p$
      used for $\hat{\P}$ and for $\P$ in~(\ref{Kuiper2}).
      Ideally, $V_1$ should be large, substantially greater than 1,
      to signal the difference between the common distribution
      of each of $X_1$,~$X_2$, \dots, $X_{n-1}$,~$X_n$ and the distribution
      with the probability density function $p$ used for $\hat{\P}$
      and for $\P$ in~(\ref{Kuiper2}).
      The numbers in parentheses in the tables indicate the order of magnitude
      of the significance level for rejecting the null hypothesis,
      that is, for asserting that the draws
      $X_1$,~$X_2$, \dots, $X_{n-1}$,~$X_n$ do not arise from $p$.
      We used~\cite{stephens2} to estimate the significance level;
      this estimate can be conservative for $V$.
\item $W_0$ is the statistic $W$ defined in~(\ref{simple}),
      with the $X_1$,~$X_2$, \dots, $X_{n-1}$,~$X_n$ in~(\ref{simple})
      drawn from a distribution with the same probability density function $p$
      and distribution function $\P$ in~(\ref{simple}).
      Ideally, $W_0$ should not be much less than 1.
\item $W_1$ is the statistic $W$ defined in~(\ref{simple}),
      with the $X_1$,~$X_2$, \dots, $X_{n-1}$,~$X_n$ in~(\ref{simple})
      drawn from a distribution that is different from the distribution
      with the probability density function $p$ used in~(\ref{simple})
      ($p$ is used both directly
      and for defining the distribution function $\P$ in~(\ref{simple})).
      Ideally, $W_1$ should be small, substantially less than 1,
      to signal the difference between the common distribution
      of each of $X_1$,~$X_2$, \dots, $X_{n-1}$,~$X_n$ and the distribution
      with the probability density function $p$ used in~(\ref{simple}).
      $W_1$ itself is the significance level
      for rejecting the null hypothesis, {\it i.e.}, for asserting that
      the draws do not arise from $p$.
\end{itemize}

\subsection{A sawtooth wave}

The probability density function $p$ for our first example is
\begin{equation}
\label{sawtooth}
p(x) = \left\{ \begin{array}{rl}
 2 \cdot 10^{-3} \cdot (x-k), & x \in (k,k+1) \hbox{ for some }
                                k \in \{0,1,\dots,998,999\} \\
                    0, & \hbox{otherwise}
               \end{array} \right.
\end{equation}
for any $x \in \R$.
The corresponding cumulative distribution function $P$ defined
in~(\ref{cumulative}) is
\begin{equation}
P(x) = \left\{ \begin{array}{rl}
 10^{-3} \cdot (x-k)^2 + 10^{-3} \cdot k, & x \in [k,k+1] \hbox{ for some }
                                            k \in \{0,1,\dots,998,999\} \\
                                         0, & x \le 0 \\
                                         1, & x \ge 1000
               \end{array} \right.
\end{equation}
for any $x \in \R$.
The distribution function $\P$ defined in~(\ref{rearranged}) is
\begin{equation}
\P(x) = \left\{ \begin{array}{rl}
 10^6 \cdot x^2/4, & x \in [0,2 \cdot 10^{-3}] \\
                1, & x \ge 2 \cdot 10^{-3}
                \end{array} \right.
\end{equation}
for any nonnegative real number $x$.

We compute the statistics $U$, $V$, and $W$ defined in~(\ref{Kuiper1}),
(\ref{Kuiper2}), and~(\ref{simple}) for two sets of i.i.d.\ draws,
first for i.i.d.\ draws distributed according to $p$ defined
in~(\ref{sawtooth}),
and then for i.i.d.\ draws from the uniform distribution on $(0,1000)$.
Table~\ref{saw_tab} displays numerical results.

For this example, the classical Kuiper statistic $U$ is unable to signal
that the draws from the uniform distribution do not arise
from $p$ defined in~(\ref{sawtooth}) for $n \le 10^7$,
at least not nearly as well as the modified Kuiper statistic $V$,
which signals the discrepancy with very high confidence for $n \ge 10^3$.
The statistic $W$ signals the discrepancy with high confidence
for $n \ge 10^3$, too.

\begin{table}
\caption{A sawtooth wave}
\label{saw_tab}
\vspace{1em}
\begin{tabular*}{\columnwidth}{@{\extracolsep{\fill}}cccclcc}
   $n$ & $U_0$ & $U_1$ & $V_0$ & \;\,\ $V_1$ & $W_0$ &   $W_1$ \\\hline
$10^1$ & .13E1 & .12E1 & .11E1 & .14E1 & .24E1 & .49E--2 \\\hline
$10^2$ & .12E1 & .18E1 & .10E1 & .21E1 & .37E0 & .45E--1 \\\hline
$10^3$ & .82E0 & .79E0 & .13E1 & .81E1 ($10^{-54}$) & .18E1 & .10E--2 \\\hline
$10^4$ & .12E1 & .17E1 & .13E1 & .25E2 ($10^{\rm -7E2}$) & .30E1 & .72E--4 \\\hline
$10^5$ & .10E1 & .12E1 & .18E1 & .79E2 ($10^{\rm -7E3}$) & .18E0 & .34E--4 \\\hline
$10^6$ & .81E0 & .14E1 & .12E1 & .25E3 ($10^{\rm -7E4}$) & .11E1 & .11E--4 \\\hline
$10^7$ & .15E1 & .19E1 & .18E1 & .79E3 ($10^{\rm -7E5}$) & .13E1 & .38E--8 \\\hline
\end{tabular*}
\vspace{1em}
\end{table}

\subsection{A step function}

The probability density function $p$ for our second example is a step function
(a function which is constant on each interval in a particular partition
of the real line into finitely many intervals).
In particular, we define
\begin{equation}
\label{high_square}
p(x) = \left\{ \begin{array}{rl}
 10^{-3}, & x \in (2k-1,2k) \hbox{ for some } k \in \{1,2,\dots,998,999\} \\
 10^{-6}, & x \in (2k,2k+1) \hbox{ for some } k \in \{0,1,2,\dots,998,999\} \\
       0, & \hbox{otherwise}
               \end{array} \right.
\end{equation}
for any $x \in \R$.
The corresponding cumulative distribution function $P$ defined
in~(\ref{cumulative}) is
\begin{equation}
P(x) = \left\{ \begin{array}{rl}
               10^{-6} \cdot k + 10^{-3} \cdot (x-k), &
               x \in [2k-1,2k] \hbox{ for some } k \in \{1,2,\dots,998,999\} \\
               10^{-6} \cdot (x-k) + 10^{-3} \cdot k, &
               x \in [2k,2k+1] \hbox{ for some } k \in \{0,1,2,\dots,998,999\}\\
               0, & x \le 0 \\
               1, & x \ge 1999
               \end{array} \right.
\end{equation}
for any $x \in \R$.
The distribution function $\P$ defined in~(\ref{rearranged}) is
\begin{equation}
\P(x) = \left\{ \begin{array}{rl}
                0, & x < 10^{-6} \\
                10^{-3}, & x \in [10^{-6},10^{-3}) \\
                1, & x \ge 10^{-3}
                \end{array} \right.
\end{equation}
for any nonnegative real number $x$.

We compute the statistics $U$, $V$, and $W$ defined in~(\ref{Kuiper1}),
(\ref{Kuiper2}), and~(\ref{simple}) for two sets of i.i.d.\ draws,
first for i.i.d.\ draws distributed according to $p$ defined
in~(\ref{high_square}),
and then for i.i.d.\ draws from the uniform distribution on $(0,1999)$.
Table~\ref{step_tab} displays numerical results.

For this example, the classical Kuiper statistic $U$ is unable to signal
that the draws from the uniform distribution do not arise
from $p$ defined in~(\ref{high_square}) for $n \le 10^6$,
at least not nearly as well as the modified Kuiper statistic $V$,
which signals the discrepancy with high confidence for $n \ge 10^2$.
The statistic $W$ does not signal the discrepancy for this example.

\begin{table}
\caption{A step function}
\label{step_tab}
\vspace{1em}
\begin{tabular*}{\columnwidth}{@{\extracolsep{\fill}}cclclcc}
   $n$ & $U_0$ & \;\,\ $U_1$ &   $V_0$ & \;\,\ $V_1$ & $W_0$ &   $W_1$ \\\hline
$10^1$ & .11E1 & .12E1 & .32E--2 & .13E1 & .10E2 &   .01E0 \\\hline
$10^2$ & .11E1 & .18E1 & .10E--1 & .46E1 ($10^{-16}$) & .10E3 &   .10E0 \\\hline
$10^3$ & .10E1 & .81E0 & .32E--1 & .16E2 ($10^{\rm -2E2}$) & .10E1 &   .10E1 \\\hline
$10^4$ & .15E1 & .17E1 & .10E--1 & .50E2 ($10^{\rm -3E3}$) & .10E2 &   .10E2 \\\hline
$10^5$ & .11E1 & .12E1 & .22E--1 & .16E3 ($10^{\rm -3E4}$) & .10E3 &   .10E3 \\\hline
$10^6$ & .70E0 & .15E1 & .19E--1 & .50E3 ($10^{\rm -3E5}$) & .10E4 &   .10E4 \\\hline
$10^7$ & .65E0 & .33E1 ($10^{-8}$) & .12E--1 & .16E4 ($10^{\rm -3E6}$) & .10E5 &   .10E5 \\\hline
\end{tabular*}
\vspace{1em}
\end{table}

\subsection{Another step function}

The probability density function $p$ for our third example is a step function
(a function which is constant on each interval in a particular partition
of the real line into finitely many intervals).
In particular, we define
\begin{equation}
\label{low_square}
p(x) = \left\{ \begin{array}{rl}
 1/10, & x \in (2k,2k+1) \hbox{ for some } k \in \{0,1,\dots,8,9\} \\
    0, & \hbox{otherwise}
               \end{array} \right.
\end{equation}
for any $x \in \R$.
The corresponding cumulative distribution function $P$ defined
in~(\ref{cumulative}) is
\begin{equation}
P(x) = \left\{ \begin{array}{rl}
               (x-k)/10, &
               x \in [2k,2k+1] \hbox{ for some } k \in \{0,1,\dots,8,9\} \\
               (k+1)/10, &
               x \in [2k+1,2k+2] \hbox{ for some } k \in \{0,1,\dots,8,9\} \\
               0, & x \le 0 \\
               1, & x \ge 19
               \end{array} \right.
\end{equation}
for any $x \in \R$.
The distribution function $\P$ defined in~(\ref{rearranged}) is
\begin{equation}
\P(x) = \left\{ \begin{array}{rl}
                0, & x < 1/10 \\
                1, & x \ge 1/10
                \end{array} \right.
\end{equation}
for any nonnegative real number $x$.

We compute the statistics $U$, $V$, and $W$ defined in~(\ref{Kuiper1}),
(\ref{Kuiper2}), and~(\ref{simple}) for two sets of i.i.d.\ draws,
first for i.i.d.\ draws distributed according to $p$ defined
in~(\ref{low_square}),
and then for i.i.d.\ draws from the uniform distribution on $(0,19)$.
Table~\ref{step2_tab} displays numerical results.

For this example, the classical Kuiper statistic $U$ signals
that the draws from the uniform distribution do not arise
from $p$ defined in~(\ref{low_square}) for $n \ge 10^3$,
but not nearly as well as the modified Kuiper statistic $V$,
which signals the discrepancy with high confidence for $n \ge 10^2$.
For this experiment, the statistic $W$ signals the discrepancy
with perfect 100\% confidence for all numbers $n$ in the table.

\begin{table}
\caption{Another step function}
\label{step2_tab}
\vspace{1em}
\begin{tabular*}{\columnwidth}{@{\extracolsep{\fill}}cclclcc}
   $n$ & $U_0$ & \;\,\ $U_1$ & $V_0$ & \;\,\ $V_1$ & $W_0$ & $W_1$ \\\hline
$10^1$ & .14E1 & .11E1 & .00E0 & .19E1 & .10E2 & .00E0 \\\hline
$10^2$ & .10E1 & .19E1 & .00E0 & .51E1 ($10^{-21}$) & .10E3 & .00E0 \\\hline
$10^3$ & .10E1 & .29E1 ($10^{-6}$) & .00E0 & .15E2 ($10^{\rm -2E2}$) & .10E4 & .00E0 \\\hline
$10^4$ & .14E1 & .95E1 ($10^{-76}$) & .00E0 & .46E2 ($10^{\rm -2E3}$) & .10E5 & .00E0 \\\hline
$10^5$ & .14E1 & .31E2 ($10^{\rm -1E3}$) & .00E0 & .15E3 ($10^{\rm -2E4}$) & .10E6 & .00E0 \\\hline
$10^6$ & .81E0 & .95E2 ($10^{\rm -1E4}$) & .00E0 & .47E3 ($10^{\rm -2E5}$) & .10E7 & .00E0 \\\hline
$10^7$ & .11E1 & .30E3 ($10^{\rm -1E5}$) & .00E0 & .15E4 ($10^{\rm -2E6}$) & .10E8 & .00E0 \\\hline
\end{tabular*}
\vspace{1em}
\end{table}

\subsection{A bimodal distribution}

\begin{figure}
\caption{The bimodal probability density function defined in~(\ref{bimodal})}
\label{bimodal_fig}
\begin{center}
\scalebox{0.5}{\includegraphics{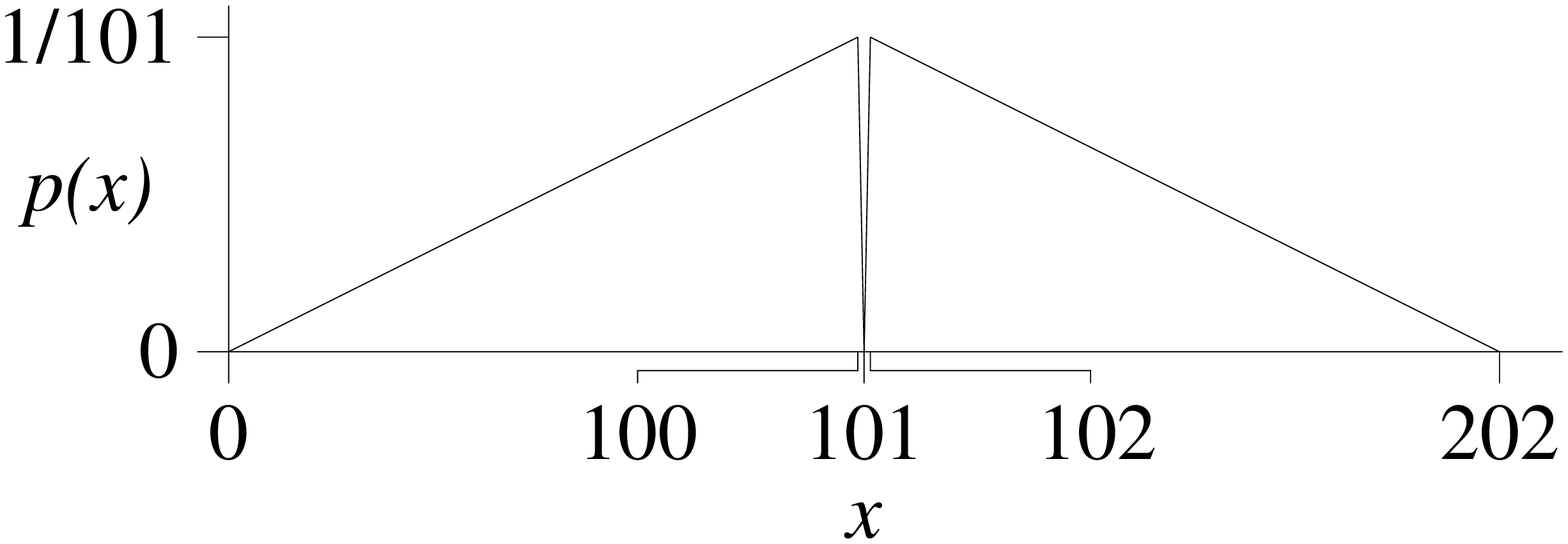}}
\end{center}
\end{figure}

The probability density function $p$ for our fourth example is
\begin{equation}
\label{bimodal}
p(x) = \left\{ \begin{array}{rl}
       x/10100, & x \in [0,100] \\
   (101-x)/101, & x \in [100,101] \\
   (x-101)/101, & x \in [101,102] \\
 (202-x)/10100, & x \in [102,202] \\
             0, & \hbox{otherwise}
               \end{array} \right.
\end{equation}
for any $x \in \R$.
Figure~\ref{bimodal_fig} plots $p$.
The corresponding cumulative distribution function $P$
defined in~(\ref{cumulative}) is
\begin{equation}
P(x) = \left\{ \begin{array}{rl}
                    x^2/20200, & x \in [0,100] \\
   (-10100 + 202 x - x^2)/202, & x \in [100,101] \\
    (10302 - 202 x + x^2)/202, & x \in [101,102] \\
 (-20604 + 404 x - x^2)/20200, & x \in [102,202] \\
                            0, & x \le 0 \\
                            1, & x \ge 202
               \end{array} \right.
\end{equation}
for any $x \in \R$.
The distribution function $\P$ defined in~(\ref{rearranged}) is
\begin{equation}
\P(x) = \left\{ \begin{array}{rl}
 (101 x)^2, & x \in [0,1/101] \\
         1, & x \ge 1/101
                \end{array} \right.
\end{equation}
for any nonnegative real number $x$.

\begin{figure}
\caption{The unimodal probability density function defined in~(\ref{unimodal})}
\label{unimodal_fig}
\begin{center}
\scalebox{0.5}{\includegraphics{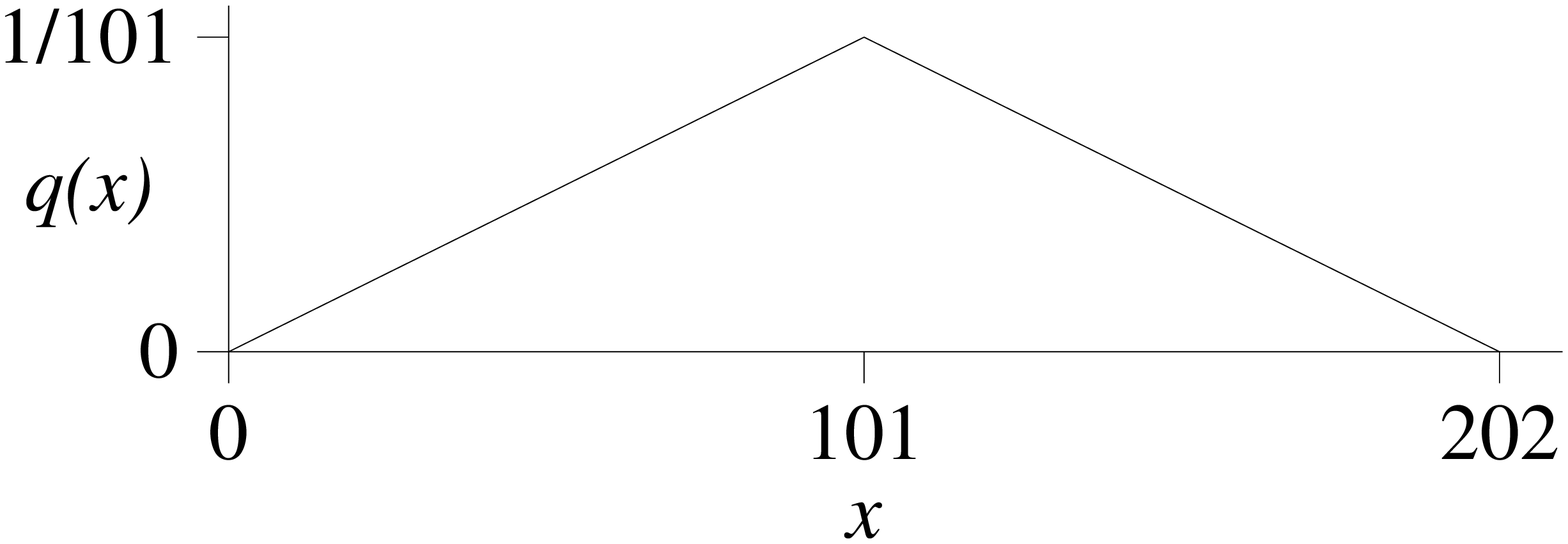}}
\end{center}
\end{figure}

We compute the statistics $U$, $V$, and $W$ defined in~(\ref{Kuiper1}),
(\ref{Kuiper2}), and~(\ref{simple}) for two sets of i.i.d.\ draws,
first for i.i.d.\ draws distributed according to $p$ defined
in~(\ref{bimodal}),
and then for i.i.d.\ draws distributed according
to the probability density function $q$ defined via the formula
\begin{equation}
\label{unimodal}
q(x) = \left\{ \begin{array}{rl}
       x/101^2, & x \in [0,101] \\
 (202-x)/101^2, & x \in [101,202]
               \end{array} \right.
\end{equation}
for any $x \in \R$.
Figure~\ref{unimodal_fig} plots $q$.
Table~\ref{bi_tab} displays numerical results.

For this example, the classical Kuiper statistic $U$ signals
that the draws from $q$ defined in~(\ref{unimodal}) do not arise
from $p$ defined in~(\ref{bimodal}) for $n \ge 10^5$,
and the modified Kuiper statistic $V$ is inferior.
The statistic $W$ signals the discrepancy with high confidence
for $n \ge 10^4$.

\begin{table}
\caption{A bimodal distribution}
\label{bi_tab}
\vspace{1em}
\begin{tabular*}{\columnwidth}{@{\extracolsep{\fill}}cclclcc}
   $n$ & $U_0$ & \;\,\ $U_1$ & $V_0$ & \;\,\ $V_1$ & $W_0$ &   $W_1$ \\\hline
$10^1$ & .11E1 & .14E1 & .11E1 & .14E1 & .11E0 & .98E--0 \\\hline
$10^2$ & .15E1 & .15E1 & .11E1 & .12E1 & .37E0 & .19E--0 \\\hline
$10^3$ & .11E1 & .10E1 & .10E1 & .13E1 & .21E1 & .70E--1 \\\hline
$10^4$ & .12E1 & .19E1 & .15E1 & .11E1 & .70E0 & .68E--3 \\\hline
$10^5$ & .10E1 & .33E1 ($10^{-8}$) & .11E1 & .18E1 & .88E0 & .40E--3 \\\hline
$10^6$ & .65E0 & .99E1 ($10^{-82}$) & .68E0 & .57E1 ($10^{-25}$) & .14E0 & .25E--7 \\\hline
$10^7$ & .89E0 & .31E2 ($10^{\rm -1E3}$) & .66E0 & .16E2 ($10^{\rm -2E2}$) & .29E0 & .25E--6 \\\hline
\end{tabular*}
\vspace{1em}
\end{table}

\subsection{A differentiable density function}

\begin{figure}
\caption{The probability density function $p$ defined in~(\ref{smooth})}
\label{smooth_fig}
\begin{center}
\scalebox{0.24}{\includegraphics{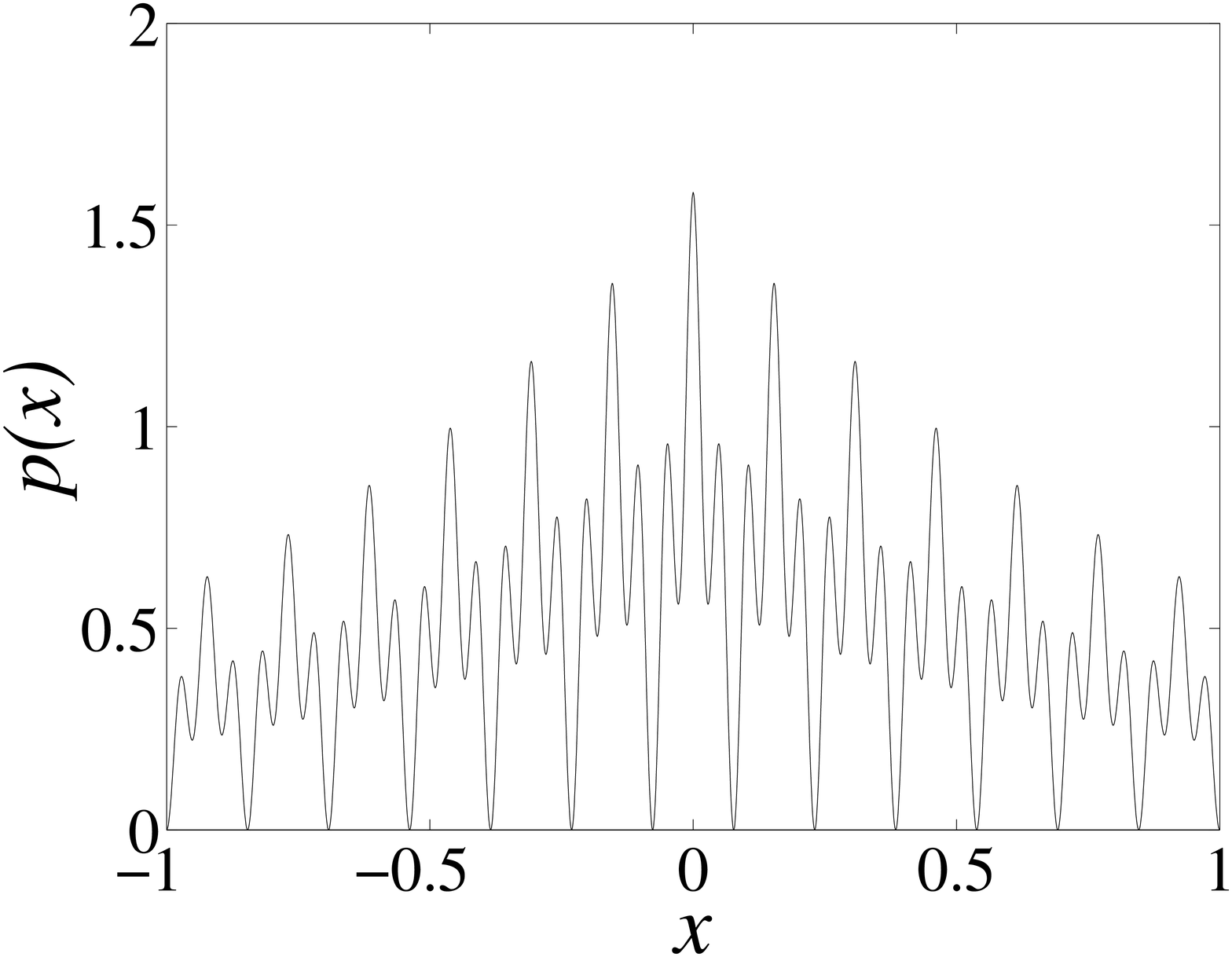}}
\end{center}
\end{figure}

\begin{figure}
\caption{The cumulative distribution function $P$ defined in~(\ref{cumulative})
         for~(\ref{smooth})}
\label{smooth_fig2}
\begin{center}
\scalebox{0.24}{\includegraphics{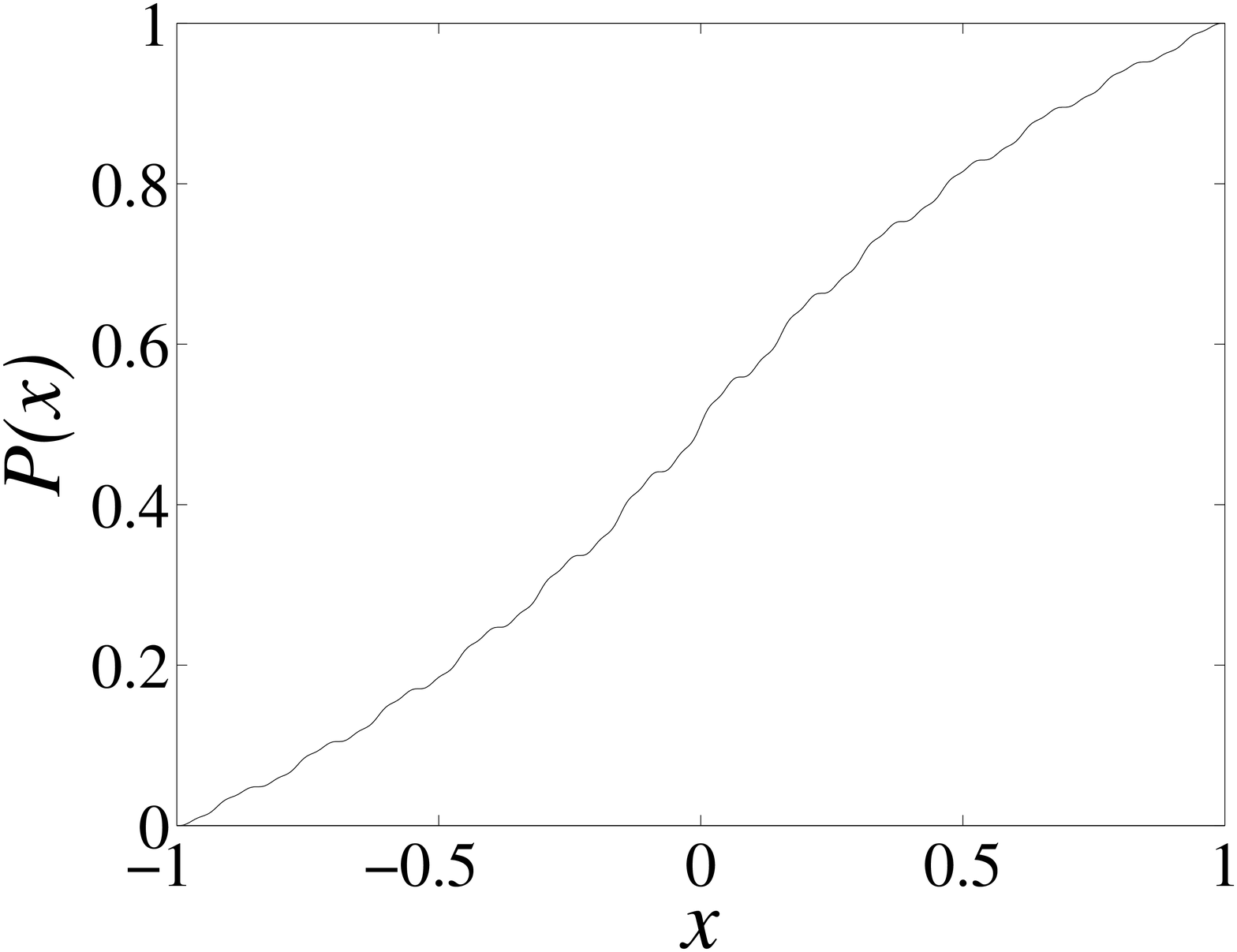}}
\end{center}
\end{figure}

\begin{figure}
\caption{The distribution function $\P$ defined in~(\ref{rearranged})
         for~(\ref{smooth})}
\label{smooth_fig3}
\begin{center}
\scalebox{0.24}{\includegraphics{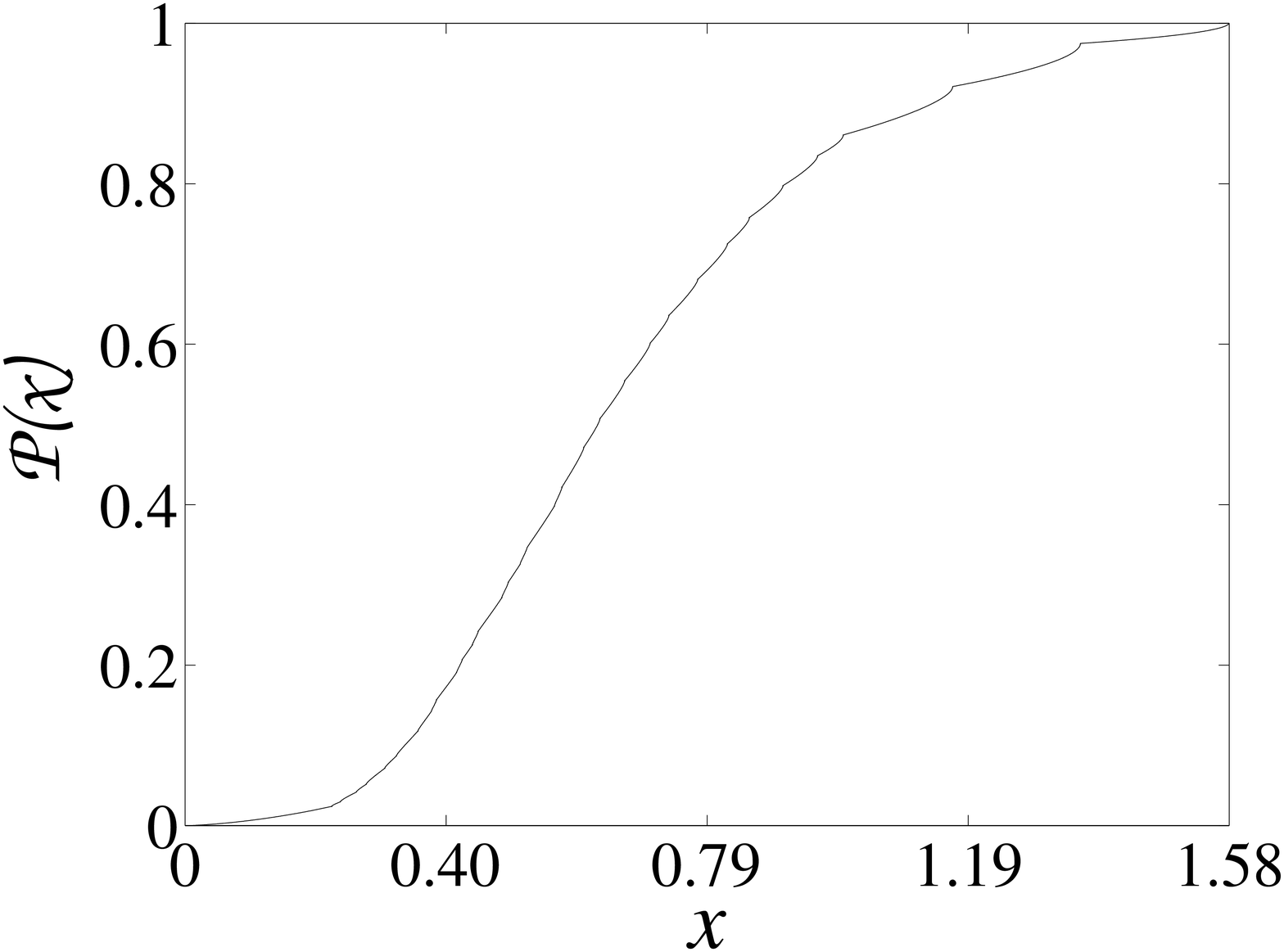}}
\end{center}
\end{figure}

The probability density function $p$ for our fifth example is
\begin{equation}
\label{smooth}
p(x) = \left\{ \begin{array}{rl}
 C \, e^{-|x|} \, (2+\cos(13 \pi x)+\cos(39 \pi x)), & x \in [-1,1] \\
                                                  0, & \hbox{otherwise}
               \end{array} \right.
\end{equation}
for any $x \in \R$, where $C \approx .4$ is the positive real number chosen
such that $\int_{-\infty}^\infty p(x) \, dx~=~1$.
Figure~\ref{smooth_fig} plots~$p$.
We evaluated numerically the corresponding cumulative distribution function $P$
defined in~(\ref{cumulative}), using the Chebfun package for Matlab
described in~\cite{chebfun}.
Figure~\ref{smooth_fig2} plots~$P$.
We evaluated the distribution function $\P$ defined in~(\ref{rearranged})
using the scheme described in the appendix below
(which is also based on Chebfun).
Figure~\ref{smooth_fig3} plots~$\P$.

We compute the statistics $U$, $V$, and $W$ defined in~(\ref{Kuiper1}),
(\ref{Kuiper2}), and~(\ref{simple}) for two sets of i.i.d.\ draws,
first for i.i.d.\ draws distributed according to $p$ defined
in~(\ref{smooth}),
and then for i.i.d.\ draws distributed according
to the probability density function $q$ defined via the formula
\begin{equation}
\label{Cauchy}
q(x) = \left\{ \begin{array}{rl}
 e^{-|x|} / (2-2\,e^{-1}), & x \in [-1,1] \\
                  0, & \hbox{otherwise}
               \end{array} \right.
\end{equation}
for any $x \in \R$.
Table~\ref{smooth_tab} displays numerical results.

For this example, the classical Kuiper statistic $U$ signals
that the draws from $q$ defined in~(\ref{Cauchy}) do not arise
from $p$ defined in~(\ref{smooth}) for $n \ge 10^4$,
but not nearly as well as the modified Kuiper statistic $V$,
which signals the discrepancy with high confidence for $n \ge 10^2$.
The statistic $W$ signals the discrepancy with high confidence
for $n \ge 10^2$, too.

\begin{table}
\caption{A differentiable density function}
\label{smooth_tab}
\vspace{1em}
\begin{tabular*}{\columnwidth}{@{\extracolsep{\fill}}cclclcc}
   $n$ & $U_0$ & \;\,\ $U_1$ & $V_0$ & \;\,\ $V_1$ & $W_0$ &   $W_1$ \\\hline
$10^1$ & .12E1 & .74E0 & .11E1 & .11E1 & .14E1 & .11E--2 \\\hline
$10^2$ & .14E1 & .11E1 & .18E1 & .30E1 ($10^{-5}$) & .13E1 & .17E--3 \\\hline
$10^3$ & .15E1 & .14E1 & .92E0 & .57E1 ($10^{-26}$) & .51E0 & .22E--4 \\\hline
$10^4$ & .86E0 & .22E1 ($10^{-3}$) & .12E1 & .16E2 ($10^{\rm -2E2}$) & .91E0 & .12E--5 \\\hline
$10^5$ & .12E1 & .58E1 ($10^{-27}$) & .12E1 & .52E2 ($10^{\rm -3E3}$) & .72E0 & .12E--6 \\\hline
\end{tabular*}
\vspace{1em}
\end{table}

\begin{remark}
For all numerical examples reported above,
at least one of the modified Kuiper statistic $V$ or the ``new'' statistic $W$
is more powerful than the classical Kuiper statistic $U$, usually strikingly so.
However, we recommend using all three statistics in conjunction,
to be conservative.
In fact, the statistics $V$ and $W$ of the present article are not able
to discern certain characteristics of probability distributions
that $U$ can, such as the symmetry of a Gaussian.
The classical Kuiper statistic $U$ should be more powerful than
its modification $V$ for any differentiable probability density function
that has only one local maximum.
For a differentiable probability density function
that has only one local maximum,
the ``new'' statistic $W$ amounts to an obvious test for outliers
--- nothing new (and far more subtle procedures for identifying outliers
are available; see, for example, \cite{simonoff1} and~\cite{davies-gather}).
Still, as the above examples illustrate,
$V$ and $W$ can be helpful with probability density functions
that have multiple local maxima.
\end{remark}

\section{Conclusions and generalizations}
\label{conclusions}

In this paper, we complemented the classical tests
of the Kolmogorov-Smirnov type with tests based on the plain fact
that it is unlikely to draw a random number whose probability is small,
provided that the draw is taken from the same distribution used
in calculating the probability
(thus, if we draw a random number whose probability is small,
then we can be confident that we did not draw the number
from the same distribution used in calculating the probability).
The numerical examples of Section~\ref{numerical} illustrate
the substantial power of the supplementary tests,
relative to the classical tests.

Needless to say, the method of the present paper generalizes
straightforwardly to probability density functions of several variables.
There are also generalizations to discrete distributions,
whose cumulative distribution functions are discontinuous.

If the probability density function $p$ involved in the definition
of the modified Kuiper test statistic $V$ in~(\ref{Kuiper2})
takes on only finitely many values,
then the confidence bounds of~\cite{stephens1}, \cite{stephens2},
and Sections~14.3.3 and~14.3.4 of~\cite{press-teukolsky-vetterling-flannery}
are conservative, yielding lower than possible confidence levels
that i.i.d.\ draws $X_1$,~$X_2$, \dots, $X_{n-1}$,~$X_n$
do not arise from $p$.
It is probably feasible to compute the tightest possible confidence levels
(maybe without resorting to the obvious Monte Carlo method),
though we may want to replace $V$ with a better statistic
when $p$ takes on only finitely many values;
for example, when $p$ takes on only finitely many values,
we can literally and explicitly rearrange $p$ to be nondecreasing
on the shortest interval outside which it vanishes,
and use the Kolmogorov-Smirnov approach on the rearranged $p$.

Even so, the confidence bounds of~\cite{stephens1}, \cite{stephens2},
and Sections~14.3.3 and~14.3.4 of~\cite{press-teukolsky-vetterling-flannery}
for the modified Kuiper test statistic $V$ in~(\ref{Kuiper2})
are sharp for many probability density functions $p$.
For example, the bounds are sharp if, for every nonnegative real number $y$,
the probability is~0 that $p(X) = y$, where $X$ is a random variable
distributed according to $p$. This covers many cases of practical interest.
In general, the tests of the present article are fully usable
in their current forms, but may not yet be optimal
for certain classes of probability distributions.

\section*{Acknowledgements}

We would like to thank Andrew Barron, G\'erard Ben Arous, Peter Bickel,
Sourav Chatterjee, Leslie Greengard, Peter W. Jones, Ann B. Lee,
Vladimir Rokhlin, Jeffrey Simonoff, Larry Wasserman, and Douglas A. Wolfe.

\section*{Appendix}

In this appendix, we describe numerical methods for constructing
the distribution function $\P$ defined in~(\ref{rearranged}).
We would be surprised if our methods turn out to be ideal in any regard,
but they seem to be adequate for our purposes,
and can leverage others' software packages to ease the implementation.

The basis of our implementation is the Chebfun package for Matlab,
described in~\cite{chebfun}.
In addition to its other capabilities,
Chebfun provides tools for the representation of piecewise-smooth
real-valued functions on bounded intervals of the real line
via Chebyshev series on adaptively chosen subintervals of the domain.
Chebfun can transform such representations in myriad ways,
including forming their derivatives and indefinite integrals.
Furthermore, Chebfun can calculate many interesting characteristics
of functions represented in this way, including local and global extrema.

Suppose that $p$ is a piecewise-smooth probability density function
that has only finitely many local extrema
on the shortest interval outside which $p$ vanishes.
Then, to compute the distribution function $\P$ defined in~(\ref{rearranged})
for a representation in Chebfun of $p$,
we perform the following four steps:

\begin{enumerate}
\item[1.] Locate the local extrema of $p$ on its computational domain
(its computational domain being the shortest closed interval outside which
$p$ vanishes).
\item[2.] Partition the computational domain of $p$ into disjoint subintervals
whose endpoints are the local extrema of $p$;
on each such subinterval, $p$ is either nondecreasing or nonincreasing.
\item[3.] On each subinterval from Step~2, form the indefinite integral of $p$,
using the subinterval's endpoint at which $p$ is smaller
for the lower limit of integration.
\item[4.] Compute a representation in Chebfun of the function $\P(x)$
given by summing up the absolute values of the indefinite integrals
from Step~3, evaluating the indefinite integrals at the points $y$
where $p(y) = x$;
if $x$ is greater than the greatest value of $p$ on a subinterval
from Step~2, then add in the greatest absolute value
of the indefinite integral on the subinterval,
while if $x$ is less than the least value of $p$ on a subinterval
from Step~2, then add in the least absolute value of the indefinite integral
(namely~$0$). On each subinterval from Step~2 for which there exists
a point $y$ such that $p(y) = x$, compute the point $y$
via bisection, trying the Newton method after 10 bisections
(and reverting to bisection if the Newton method fails to produce accuracy
of a digit less than the machine precision after 5 Newton steps).
\end{enumerate}
(Step~4 describes a procedure for evaluating $\P(x)$ at an arbitrary point $x$.
Given this procedure, Chebfun automates the construction
of a highly accurate representation of $\P$ that can be evaluated efficiently
at arbitrary points.)


\bibliographystyle{siam}
\bibliography{stat}

\begin{thebibliography}{10}

\bibitem{barron}
{\sc A.~R. Barron}, {\em Uniformly powerful goodness of fit tests}, Ann.
  Statist., 17 (1989), pp.~107--124.

\bibitem{bickel-ritov-stoker}
{\sc P.~J. Bickel, Y.~Ritov, and T.~M. Stoker}, {\em Tailor-made tests for
  goodness of fit to semiparametric hypotheses}, Ann. Statist., 34 (2006),
  pp.~721--741.

\bibitem{bickel-rosenblatt}
{\sc P.~J. Bickel and M.~Rosenblatt}, {\em On some global measures of the
  deviations of density function estimates}, Ann. Statist., 1 (1973),
  pp.~1071--1095.

\bibitem{davies-gather}
{\sc L.~Davies and U.~Gather}, {\em The identification of multiple outliers},
  J. Amer. Statist. Assoc., 88 (1993), pp.~782--792.

\bibitem{fan}
{\sc J.~Fan}, {\em Test of significance based on wavelet thresholding and
  {N}eyman's truncation}, J. Amer. Statist. Soc., 91 (1996), pp.~674--688.

\bibitem{hollander-wolfe}
{\sc M.~Hollander and D.~A. Wolfe}, {\em Nonparametric Statistical Methods},
  Wiley, New York, second~ed., 1999.

\bibitem{inglot-ledwina}
{\sc T.~Inglot and T.~Ledwina}, {\em Asymptotic optimality of data-driven
  {N}eyman's tests for uniformity}, Ann. Statist., 24 (1996), pp.~1982--2019.

\bibitem{khamis}
{\sc H.~J. Khamis}, {\em The two-stage $\delta$-corrected
  {K}olmogorov-{S}mirnov test}, J. Appl. Statist., 27 (2000), pp.~439--450.

\bibitem{press-teukolsky-vetterling-flannery}
{\sc W.~Press, S.~Teukolsky, W.~Vetterling, and B.~Flannery}, {\em Numerical
  Recipes}, Cambridge University Press, Cambridge, UK, third~ed., 2007.

\bibitem{rayner-thas-best}
{\sc J.~C.~W. Rayner, O.~Thas, and D.~J. Best}, {\em Smooth Tests of Goodness
  of Fit}, Wiley, second~ed., 2009.

\bibitem{reschenhofer}
{\sc E.~Reschenhofer}, {\em Combining generalized {K}olmogorov-{S}mirnov
  tests}, InterStat, June 2008 (2008), pp.~1--15.

\bibitem{simonoff1}
{\sc J.~S. Simonoff}, {\em Outlier detection and robust estimation of scale},
  J. Stat. Comput. Simul., 27 (1987), pp.~79--92.

\bibitem{simonoff2}
\leavevmode\vrule height 2pt depth -1.6pt width 23pt, {\em Smoothing Methods in
  Statistics}, Springer-Verlag, New York, 1996.

\bibitem{stein-weiss}
{\sc E.~M. Stein and G.~Weiss}, {\em Introduction to {F}ourier Analysis on
  {E}uclidean Spaces}, Princeton University Press, Princeton, NJ, 1971.

\bibitem{stephens1}
{\sc M.~A. Stephens}, {\em The goodness-of-fit statistic {$V_n$}:
  {D}istribution and significance points}, Biometrika, 52 (1965), pp.~309--321.

\bibitem{stephens2}
\leavevmode\vrule height 2pt depth -1.6pt width 23pt, {\em Use of the
  {K}olmogorov-{S}mirnov, {C}ramer--{V}on-{M}ises and related statistics
  without extensive tables}, J. Roy. Statist. Soc. Ser. B, 32 (1970),
  pp.~115--122.

\bibitem{chebfun}
{\sc L.~N. Trefethen, N.~Hale, R.~B. Platte, T.~A. Driscoll, and R.~Pach\'on},
  {\em Chebfun, version 3}.
\newblock http://www.maths.ox.ac.uk/chebfun, Oxford University, 2009.

\bibitem{wasserman}
{\sc L.~Wasserman}, {\em All of Statistics}, Springer, 2003.

\end{thebibliography}

\end{document}